\documentclass[12pt]{article}
\usepackage{mathrsfs}
\usepackage{graphicx}
\usepackage{amssymb,amsmath,amsthm}
\usepackage{epstopdf}
\usepackage{tocbibind}
 \textwidth = 6.5 in \textheight = 9 in \oddsidemargin =
0.0 in \evensidemargin = 0.0 in \topmargin = 0.0 in \headheight =
0.0 in \headsep = 0.0 in
\def\ba{\begin{eqnarray}}
\def\ea{\end{eqnarray}}

\def\:{\boldsymbol{:}}

\begin{document}
\begin{flushright}
\end{flushright}
\begin{center} 
\vglue .06in
{\Large \bf {Propagation Constraints and Classical Solutions in K-essence Like Theories}}\\[.5in]

{\bf Ratindranath Akhoury and Christopher S. Gauthier}\\
[.1in]
{\it{Michigan Center for Theoretical Physics\\
Randall Laboratory of Physics\\
University of Michigan\\
Ann Arbor, Michigan 48109-1120, USA}}\\[.2in]
\end{center}
\begin{abstract}
\begin{quotation}
We consider two examples of solutions of the equations of motion of scalar field theories with higher derivatives. These are the cosmology of the rolling tachyon and static spherically symmetric solutions of the scalar field in flat space. By requiring that the field equations always be hyperbolic and that the speed of propagation of the small fluctuations are not superluminal, we find constraints on the form of the allowed interactions in the first case and on the choice of boundary conditions in the latter. For the rolling tachyon we find a general class of models which have the property that at large times the tachyon matter behaves essentially like a non-relativistic gas of dust. 
For the spherically symmetric solutions we show how causality influences the choice of boundary conditions and those which are finite at the origin are shown to have negative energy density there.
\end{quotation}
\end{abstract}
\newpage
\section{Introduction}  Scalar theories with higher derivatives play an essential role in the effective
field theory approach (for reviews see \cite{eftreview}). An example is provided by the chiral lagrangian which provides a good description of the strong dynamics at low energies. Applications of higher derivative theories to cosmology have also become popular in the last few years: examples here are effective field theories of the rolling tachyon \cite{tachyon}, DBI inflation \cite{dbi}, and k-essence \cite{kessence} which attempts to provide a dynamical explanation of  the so called coincidence problem and the accelerated expansion of the universe. Recently \cite{shiu}, such higher derivative actions have been shown to enhance the non-gaussian fluctuations in the cosmic microwave background. Theories with higher derivatives have the possibility of modifying the dispersion relations and hence may potentially lead to superluminal propagation. This aspect has been studied in detail in \cite{arkanihamed} where it was shown that causality and the absence of superluminal propagation require certain coefficients of the effective lagrangian to be positive definite which in turn has consequences for phenomenology \cite{arkanihamed}, \cite{ira}. Thus the constraints of causality and hyperbolicity of the equations of motion play a particularly important role in such theories. Another recent striking example is the no-go theorem proved in \cite{durer}. Here it was shown that in the context of the original k-essence theories \cite{kessence}, it is impossible to simultaneously resolve the coincidence problem and the accelerated expansion of the universe without violating causality.

In this paper we apply the constraints \cite{suskind, wald, bruneton, dmhalos} that the equation of motion of the scalar field has a well defined initial value problem and there is no superluminal propagation of the small fluctuations around classical solutions in higher derivative theories. In particular we discuss in sections 3 and 4 respectively the case of the rolling tachyon and the static solutions to the equations of motion of a general scalar theory with higher derivatives. For the case of the tachyon we consider a general lagrangian of the form $L = V(\phi) K(X)$, with $V$ the potential for the tachyon and $X = g^{\mu\nu}\partial_{\mu}\phi\partial_{\nu}\phi$ and find the constraints on $K(X)$ and the potential such that the the energy density is finite but the equation of state parameter goes to zero at large times up to  small corrections. We find that in order achieve this it is not nescessary  for $K(X)$ to vanish as $\dot{\phi} \rightarrow 1$ but only that it be bounded. Other constraints on $K$ are obtained which allows for a more general framework for the rolling tachyon than was previously considered. 
The only constraint on the potential is that $a^3V \rightarrow 0$ at large times, where $a$ is the scale factor. The physical motivation is that the tachyon could then be considered as a possible candidate for dark matter \cite{shiudm}. In section 4 we discuss the static spherically symmetric solutions to the equations of motion for the most general scalar field lagrangian with higher derivatives in flat space which are consistent with hyperbolicity and causality. We find the interesting result that for scalar field solutions which are finite at the origin, causality requires its first derivative to vanish there, and even though the total energy is positive, the energy density for such solutions is negative at the origin. A  physical motivation for this study arises from the possibilty that such scalar fields could describe the dark matter halos around galaxies \cite{dmhalos}. In section 2 we set up the problem and review some results concerning the criteria for superluminal propagation and hyperbolicity of the scalar field equations. In the concluding section we discuss the results.
\section{Preliminaries} In this paper we will be interested in scalar field theories with a lagrangian
of the general form $L = \frac{1}{2}F(X,\phi)$. Here, $X = \partial^{\mu}\phi \partial_{\mu}\phi$
We will first discuss the case of flat space time and at the end comment on the general case in the presence of gravity. 

The equations of motion for the scalar field are given by:
\begin{eqnarray}
G^{\mu\nu}\partial_{\mu}\partial_{\nu}\phi = \frac{1}{2}\{L_{\phi} - 2XL_{X\phi}\} \\
G^{\mu\nu} = L_X\eta^{\mu\nu} + 2L_{XX}\partial^{\mu}\phi \partial^{\nu}\phi. \label{fieldmetric}
\end{eqnarray}
Throughout this paper we will be using the notation,
\begin{equation}
L_X = {\partial L \over \partial X} ; \quad L_{\phi} = {\partial L \over \partial \phi}
\end{equation}
and so on. In (\ref{fieldmetric}), $G^{\mu\nu}$ plays the role of an effective metric in which the scalar field propagates. For an equation of this type to have a well defined initial value problem and to satisfy
global hyperbolicity, the following conditions must hold \cite{suskind,wald,bruneton}
\begin{equation}
L_X > 0, \quad L_X + 2XL_{XX} > 0 \label{ineq1}
\end{equation}
If $u=0$ is the characteristic surface and $n^{\mu} = \partial^{\mu}u$ then the speed of propagation of the small disturbances is given by solving 
\begin{equation}
L_Xn^2 + 2L_{XX}(n^{\mu}\partial_{\mu}\phi)^2 = 0.
\end{equation}
From this, one deduces the maximum speed to be 
\begin{equation}
{n^0 \over |\vec{n}|} = {W_0\left({\vec{n}. \vec{W} \over |\vec{n}|}\right) + \sqrt{ 1 + W_0^2 - {\left({\vec{n}.\vec{W} \over |\vec{n}|}\right)}^2} \over 1 + W_0^2}
\end{equation}
where, $W_{\mu} = \sqrt{2L_{XX} \over L_X}\partial_{\mu}\phi$.
The two cases discussed in this paper are the time-like spatially homogenous and static  spherically symmetric ones. The expressions for the propagation speeds in the two cases are respectively,
\begin{eqnarray}
{n^0 \over |\vec{n}|} = \{{ L_X \over L_X + 2XL_{XX}}\}^{1 \over 2}, \quad X = \dot{\phi}^2 \\
{n^0 \over |\vec{n}|} = \{{L_X + 2XL_{XX} \over L_X}\}^{1 \over 2}, \quad X = -\phi'^2.
\end{eqnarray}
From these it is easy to see that there is superluminal propagation when $L_{XX} < 0$. In summary, the conditions of hyperbolicity and no superluminal propagation may be stated as:
\begin{equation}
L_X > 0, \quad L_X + 2XL_{XX} > 0, \quad L_{XX} \geq 0.          \label{ineq1}
\end{equation}
For future reference we note that in the static spherically symmetric case when there is no superluminal propagation,
\begin{equation}
{ L_X \over L_X + 2XL_{XX} } \geq 1. \label {speed} 
\end{equation}

When gravity is included, the effective scalar metric now becomes:
\begin{equation}
G^{\mu\nu} = L_Xg^{\mu\nu} + 2L_{XX}\partial^{\mu}\phi \partial^{\nu}\phi \label{metric2}
\end{equation}
 We require this metric to be Lorentzian. In particular in order to have a consistent definition of temporal and spatial directions the largest eigenvalue of (\ref{metric2}) must be positive while the other three must be negative. This can be shown to be true \cite{wald,bruneton} only if the first two conditions in (\ref{ineq1}) are satisfied while the last one once again avoids \cite{suskind, arkanihamed} superluminal propagation. 
\section{The Rolling Tachyon} Sen \cite{tachyon} has discussed  the qualitative dynamics of a tachyon field in the background of an unstable D-brane system and conjectured that the simplest description within an effective field theory framework can be provided by the following lagrangian, 
$L = V(\phi)K(X)$ with, $\phi$ the scalar field (dot denotes derivative with respect to $t$) and,
\begin{eqnarray}
K(x) = - \sqrt{1- {\dot{\phi}}^2} \nonumber \\
V(\phi) = V_0 \exp(-\phi). \label{tachyonsen}
\end{eqnarray}
The cosmology of this model in the FRW background has been studied in \cite{tachyon}, \cite{shiudm},
and a particularly surprising result is the existence of solutions with exponentially vanishing pressure at large times but a non-zero energy density.  Since there is no compelling reason for the precise forms Eq.(\ref{tachyonsen}), in this section we keep $K$ and $V$ arbitrary (apart from the mild assumptions below) and determine from the constraints of causality the conditions under which the equation of state for tachyonic matter becomes $\omega = 0$ at large times up to small corrections. The tachyon could then be considered a dark matter candidate in a wider class of models than originally envisioned. 

Consistent with the fact that we are dealing with the case of a rolling tachyon, we will make the following assumptions about the potential $V$ and the kinetic term $K$:  (1) $K \leq 0$ (2) The range of $\dot{\phi}$ is bounded. We will take the upper limit of $\dot{\phi}$ to be $1$ in appropriate units. (3) K is bounded as $\dot{\phi} \rightarrow 1$ (4) The potential $V(\phi)$  is positive, has a maximum at $\phi = 0$ and monotonically decreases to zero at $\phi = \infty $ at large times where it is a minimum. 

The equations of motion for the scalar field and the scale factor $a(t)$ are in units $\frac{8\pi G}{3} = 1$: 
\begin{eqnarray}
\ddot{\phi} = -3H{L_X \over L_X + 2XL_{XX}}\dot{\phi} - \frac{1}{2}{\frac{\partial \epsilon_t }{\partial \phi} \over  L_X + 2XL_{XX}} \nonumber \\
H^2 = \rho = \epsilon_t + \epsilon_m. \label{cosmodynamics}
\end{eqnarray}
For the homogenous FRW background $X = {\dot{\phi}}^2 > 0$.
$\epsilon_t = 2XL_X - L$ is the tachyon energy density and $\epsilon_m $ is that of the rest of matter and $\rho$ is the total energy density. $H = \frac{\dot{a}}{a}$ is the Hubble factor. Note that from the inequality Eq.(\ref{ineq1} ) and $L < 0$, it is easy to see that $\epsilon_t > 0$. Thus the nonvanishing of the energy density at all times including late times is strictly a consequence of $L_X > 0$.
The equation of state parameter for the tachyon is,
\begin{equation}
\omega_t = {L \over 2XL_X - L} = {K \over 2XK_X - K} \label{omega}
\end{equation}
Using the $\phi$ field equation of motion it is straightforward to show that 
\begin{equation}
\dot{\epsilon_t} = {d \over dt}( 2XL_X - L ) = - 6HXL_X = -3H(1 + \omega_t)\epsilon_t \label{epsilondot}
\end{equation}
The inequality in Eq. (\ref{ineq1}) then implies that the tachyon energy density is a monotonically decreasing function of time and $\omega_t > -1$.  The Hubble factor itself is monotonically decreasing in time as can be seen from 
\begin{equation}
\dot{H} = - \frac{3}{2} \left( (1+\omega_m)\epsilon_m + (1+\omega_t)\epsilon_t \right),
\end{equation}

Defining $y = \sqrt X$ and using the factorized form for the tachyon lagrangian, the tachyon equation of motion may be written as,
\begin{equation}
{dy \over dt} = - {yK_y - K \over K_{yy}}\left(3H\{{K_y \over yK_y - K}\} + {{\partial V \over \partial \phi} \over V}\right).  \label{scalareq2}
\end{equation}
The constraints given in Eq. (\ref{ineq1}) for the initial value problem to be well defined and the absence of superluminal propagation are expressed in terms of the new variable as:
\begin{equation}
K_y > 0, \quad  K_{yy} > 0,  \quad K_{yy} > {K_y \over y}. \label{ineq2}
\end{equation}
Note that whenever $V \rightarrow 0$, $K_y \rightarrow \infty$ such that $L_X > 0$. As we will see below, it is this simple fact that guarantees that the tachyon energy density is nonzero and positive in the limit $t \rightarrow \infty$, while $\omega_t$ vanishes. 
Let us consider Eq. (\ref{scalareq2}) at large times. We first discuss the conditions on the potential under which  the second term in the brackets is dominant. Let us define the term inside the curly brackets in this equation as $g$, then
\begin{equation}
{dg \over dy} = { - KK_{yy} \over (yK_y - K)^2 }.
\end{equation}
Since $K < 0$  we see from (\ref{ineq2}) that ${dg \over dy} > 0$. The maximum value of $g$ is thus at $y=1$ which is $g_{max} \leq 1$. Moreover, $H$ is monotonically decreasing. Let us now write for large times  $\phi = t + \theta(t)$ with $\theta(t) \ll t$. Then it is easy to check from the above results that the second term inside the brackets in Eq.(\ref{scalareq2}) dominates over the first for large times as long as $V \rightarrow 0$ faster than ${1 \over a^3}$ as $t \rightarrow \infty$. This condition on the potential will reappear below. Since the overall factor outside the brackets in (\ref{scalareq2}) is negative, and since ${\partial V \over \partial \phi} < 0$ from our assumptions, the above discussion shows that $y$ is monotonically increasing as it goes to $1$ at large times. In addition, since $y$ is bounded at $y=1$, $\dot{y} = 0$ at $y =1$. Therefore we conclude that as $y \rightarrow 1$,
\begin{equation}
{K_y \over K_{yy}} = 0 ;  {K \over K_{yy}} = 0 \label{kzero}
\end{equation}
As mentioned earlier, $K_y \rightarrow \infty$ for large times while $K$ is bounded. Thus the second of the above conditions is not a new requirement since the first implies that as $y \rightarrow 1$,
$K_{yy} > K_y$.
It should be noted that the condition for the absence of superluminal propagation only implies that 
\begin{equation}
{K_y \over K_{yy}} < y,
\end{equation}
Thus the condition (\ref{kzero}) is much stronger. 

Let us now expand the equation (\ref{scalareq2}) about the point $y = 1$ by writing,
$\phi = t + \theta(t)$ with $\dot{\theta} < 0$ and $\theta \ll t$. Using (\ref{kzero}), it is straightforward to get,
\begin{eqnarray}
\ddot{\theta} = - \dot{\theta}\lambda\{ 3H + {{\partial V \over \partial \phi} \over V}\} \\
\lambda = \left( 1- {K_{yyy} K_y \over K_{yy}^2} \right)(y=1). \label{lambda}
\end{eqnarray}
Integrating this we get (taking $\dot{\phi} \approx 1$ to leading order)
\begin{equation}
\dot{\theta} = - \alpha a^{-3\lambda} V^{- \lambda} .
\end{equation}
where $\alpha$ is a constant and consistency with the boundary conditions require $\lambda < 0$. 
Since $\theta \ll t$, we see that with a negative $\lambda$,  $\dot{\theta}$ vanishes like $a^3V \rightarrow 0$ as $t \rightarrow \infty$, which is exactly the condition derived earlier for the term involving the potential V to dominate over the first one in Eq. (\ref{scalareq2}). $\lambda < 0$  or equivalently at $y=1$, $K_{yyy} > K_{yy}$ is then a new constraint on the allowed forms of $K$.

We are now in a position to prove that the equation of state parameter vanishes at $y=1$ up to small
corrections. From Eq.(\ref{omega}) we can obtain,
\begin{equation}
{d\omega_t \over dy} = { yK_y^2 - KK_y - yKK_{yy} \over (yK_y - K)^2}.
\end{equation}
Since $K \leq 0$, and $K_y$ and $K_{yy}$ are both positive, we see that $\omega_t$ is a monotonically increasing function of $y$. Its maximum is therefore at $y=1$. Near $y=1$ we can write 
\begin{equation}
\omega_t \approx {K(1) + \dot{\theta}K_y(1) \over K_y(1)}.
\end{equation}
However we have argued above that $K_y(1)$ is infinite, thus $\omega_t = 0$ apart from corrections which vanish like $a^3V$ at large times.
\section{Background Static Solutions Consistent With Causality} We next consider the static spherically symmetric solutions to the equations of motion of the scalar field in flat spacetime (prime denotes the derivative with respect to $r$).
\begin{equation}
\phi'' + \frac{2}{r}\{{L_X \over L_X + 2XL_{XX}}\}\phi' + \frac{1}{2}\{{L_{\phi} - 2XL_{X\phi} \over  L_X + 2XL_{XX}}\} = 0.   \label{staticeq}
\end{equation}
In the above, $X = -\phi'^2$ and from section 1 the combined constraints of hyperbolicity and absence of superluminal propagation now give the following bound for the coefficient of the $\phi'$ term for all $r$ (see Eq. (\ref{speed})):
\begin{equation}
\delta = {L_X \over L_X + 2XL_{XX}} \geq1.
\end{equation}
Here we consider only  solutions to (\ref{staticeq}) which are finite at the origin. We first want to determine the appropriate boundary condition for $\phi'$ at $r=0$. We will use a series expansion method for $\phi$ near $r=0$ to guide us to the correct choice.
Even though the coefficient $\delta$ is not a constant but dependent on $\phi$, we know that independent of $r$, $\delta \geq 1$, so to find the indical equation, which is all we are interested in to determine the boundary condition for $\phi'$, we may treat it as such. The same applies for the last term in (\ref{staticeq}) as long as we restrict ourselves to solutions which are finite at the origin. These two complications do not affect the indical equation. With this in mind, let us look for a series solution of the form:
\begin{equation}
\phi = r^s(c_0 + c_1r + c_2r^2 + c_3r^3 + ....)
\end{equation}
From this we get the indical equation $s(s-1+2\delta) = 0$. Since causality requires $\delta \geq 1$ for any $r$ we see that for $\phi$ to be finite at the origin
only the solution with $s=0$ is allowed. Substituting this expansion into (\ref{staticeq})
we see by matching equal powers of $r$ that $c_1 =0$. Thus the boundary condition for this problem which is consistent with causality and the finiteness of $\phi$ at the origin is $\phi' =0, r=0$. We now consider the analog of Eq. (\ref{epsilondot}). Let us define $\gamma = -2XL_X + L$. Then using the equation of motion we obtain
\begin{equation}
{d\gamma \over dr} = -\frac{4}{r}\phi'^2L_X. \label{gamma}
\end{equation}
Since $L_X >0$, we see that $\gamma$ is a monotonically decreasing function of $r$.  The minimum of $\gamma$ is therefore at infinity. As $r \rightarrow \infty$, the solutions to the equations of motion must be such that $\gamma \rightarrow 0$ faster than ${1 \over r^3}$ in order to keep the total energy content finite. This implies that at $r \rightarrow 0$, $\gamma > 0$. From the boundary condition on $\phi'$ at $r=0$ we see that here, $\gamma = L > 0$. On the other hand, in the static limit, the hamiltonian density $\cal{H}$ = $- L$. Thus we conclude that at $r=0$, the energy density $\cal{H}$ is negative. It is easy to see that the total energy in the static limit is, however, always positive:
\begin{equation}
E = - 4\pi \int r^2dr L = -4\pi \int(\gamma -2\phi'^2L_X)r^2dr.
\end{equation}
Consider the integral over $\gamma$. Integrating by parts and using the fact that $\gamma$ vanishes faster than ${1 \over r^3}$ as $r$ approaches infinity, we get
\begin{equation}
\int \gamma r^2 dr = {-1 \over 3}\int{d\gamma \over dr}r^3dr = {4 \over 3}\int\phi'^2L_Xr^2dr,
\end{equation}
where the last equality follows from (\ref{gamma}). Combining everything we see that
\begin{equation}
E = {8 \pi \over 3}\int\phi'^2L_Xr^2dr,
\end{equation}
which is manifestly positive definite.

When such a theory is coupled to the Schwarzschild metric, we can look for solutions to the combined equations for both gravity and scalar matter. Such a situation could be relevant for understanding the formation of dark matter halos around galaxies \cite{dmhalos}. Though the above analysis has been performed in flat space-time, our considerations indicate that at least solutions of the scalar field equations which are finite at the origin should not be relevant to such a scenario. The detailed question of the solutions of the scalar field in the presence of gravity needs further investigation. Nevertheless it is interesting that  the model we have considered in this section has solutions which violate the weak energy condition at the origin.

\section{Conclusions} Using the requirement that the field equations are always hyperbolic (and hence the cauchy problem is well defined) we have obtained a set of consequences for two different problems of physical interest. 

For the case of the rolling tachyon in a homogenous FRW background, we have obtained constraints on the forms of the potential and the kinetic terms such that the equation of state of the tachyon vanishes at large times up to small corrections. The tachyon could then be considered a dark matter candidate. The key observation here was that what is required for this to happen is that $K$ remains bounded but not necessarily zero at large times, but $K_y$ goes to infinity. The latter in fact guarantees that the energy is non vanishing in this limit. Other requirements are given by Eqs.(\ref{kzero}), the potential $V$ is such that $a^3V \rightarrow 0$ at large times, and that $\lambda$ defined in (\ref{lambda}) be negative. It is easy to check that the choice (\ref{tachyonsen}) does in fact satisfy all the requirements, but  is not unique. The class of models is thus larger than the original.

We have also looked quite generally at the problem of the static spherically symmetric solutions to the equations of motion of the scalar field described by the lagrangian of section 1 and found that if we require the finiteness of the scalar field at the origin then the solutions consistent with causality have the property that the energy density becomes negative at the origin. This example brings out very clearly the role that causality plays in the choice of boundary conditions.
There have been attempts in the literature \cite{dmhalos}  to use such scalar field models to describe dark matter halos around galaxies. Clearly, solutions which are finite at the origin will not do this job. However it is interesting to speculate if this
negative energy density at the origin is indicative of an attractive force, analogous to the casimir effect (but of course classical), at the center of galaxies.

\section{Acknowledgements} We would like to thank Paul Federbush for discussions. This work was supported by the US department of energy.

\newpage

\end{document}